\documentclass[preprint,journal]{vgtc}       

\ifpdf
  \pdfoutput=1\relax                   
  \usepackage{graphicx}                
  \DeclareGraphicsExtensions{.pdf,.png,.jpg,.jpeg} 
\else
  \usepackage{graphicx}                
  \DeclareGraphicsExtensions{.eps}     
\fi%

\graphicspath{{figures/}{pictures/}{images/}{./}} 

\usepackage{microtype}                 
\PassOptionsToPackage{warn}{textcomp}  
\usepackage{textcomp}                  
\usepackage{mathptmx}                  
\usepackage{times}                     
\usepackage{cite}                      
\usepackage{tabu}                      
\usepackage{booktabs}                  

\usepackage{amsfonts}

\usepackage{hyperref}
\usepackage{xcolor}
\usepackage{soul}
\newcommand{\sysname}{\emph{COVID-view}}

\onlineid{1205}

\vgtccategory{Application}
\vgtcpapertype{Application}

\title{\sysname: Diagnosis of COVID-19 using Chest CT}

\author{Shreeraj Jadhav$^*$, Gaofeng Deng$^*$, Marlene Zawin, Arie E. Kaufman, \textit{Fellow, IEEE}}
\authorfooter{
\item
 Shreeraj Jadhav, Gaofeng Deng, and Arie E Kaufman are with the Department of Computer Science at Stony Brook University. E-mail: (sdjadhav, dgaofeng, ari)@cs.stonybrook.edu  ~~~~*Equal contribution
\item
 Marlene Zawin is with the Department of Radiology at Stony Brook University Hospital. E-mail: marlene.zawin@stonybrookmedicine.edu
}

\shortauthortitle{Biv \MakeLowercase{\textit{et al.}}: Global Illumination for Fun and Profit}

\abstract{
Significant work has been done towards deep learning (DL) models for automatic lung and lesion segmentation and classification of COVID-19 on chest CT data. However, comprehensive visualization systems focused on supporting the dual visual+DL diagnosis of COVID-19 are non-existent. We present \sysname, a visualization application specially tailored for radiologists to diagnose COVID-19 from chest CT data. The system incorporates a complete pipeline of automatic lungs segmentation, localization/isolation of lung abnormalities, followed by visualization, visual and DL analysis, and measurement/quantification tools. 
Our system combines the traditional 2D workflow of radiologists with newer 2D and 3D visualization techniques with DL support for a more comprehensive diagnosis. \sysname\ incorporates a novel DL model for classifying the patients into positive/negative COVID-19 cases, which acts as a reading aid for the radiologist using \sysname\, and provides the attention heatmap as an explainable DL for the model output. 
We designed and evaluated \sysname\ through suggestions, close feedback and conducting case studies of real-world patient data by expert radiologists who have substantial experience diagnosing chest CT scans for COVID-19, pulmonary embolism, and other forms of lung infections.
We present requirements and task analysis for the diagnosis of COVID-19 that motivate our design choices and results in a practical system which is capable of handling real-world patient cases. 
} 


\CCScatlist{ 
 \CCScat{K.6.1}{Management of Computing and Information Systems}%
{Project and People Management}{Life Cycle};
 \CCScat{K.7.m}{The Computing Profession}{Miscellaneous}{Ethics}
}

\teaser{
  \centering
  \includegraphics[width=\linewidth,trim={0 0 0 0},clip]{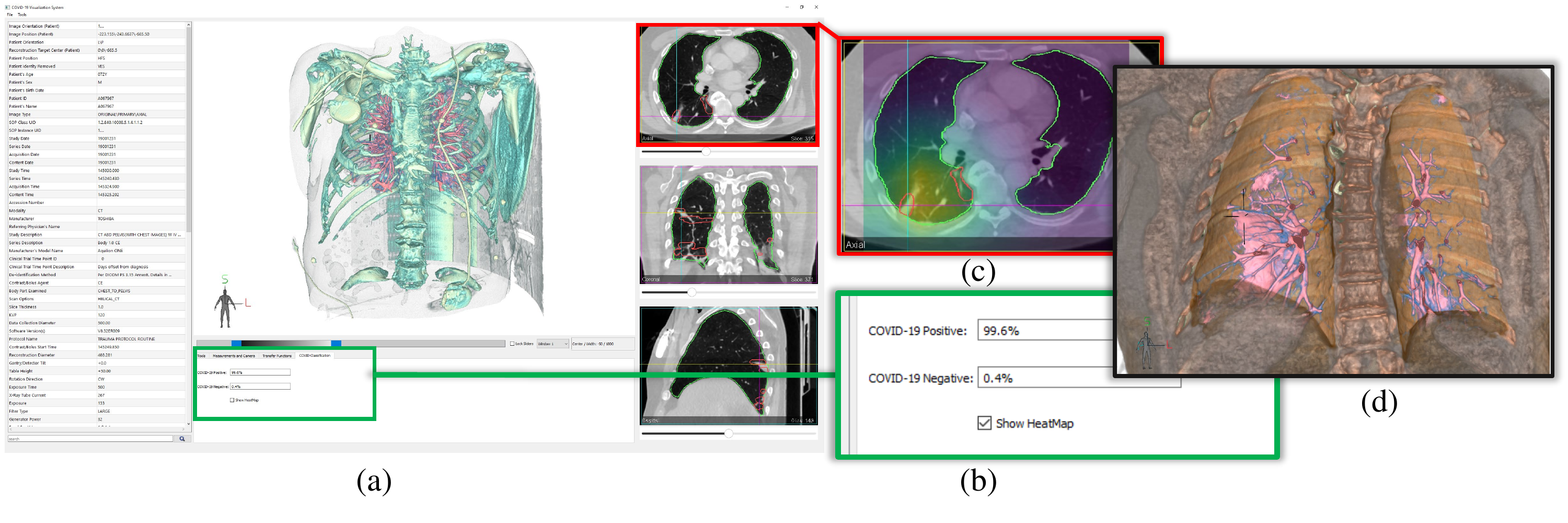}
  \caption{An overview of the user interface and visualizations in our \sysname\ system.
  (a) Snapshot of the user interface containing 2D and 3D views. (b) Results from our deep learning binary classifier showing COVID positive diagnosis. (c) Axial view with lesion segmentation outlines (red) and Grad-CAM activation heatmap overlay for explainable classification results. Both heatmap and lesion localization outlines point towards the interlobular septal thickening. (d) Clipped 3D view with lungs and vicinity context volume rendered together showing the 3D rendering of the thick interlobular septation (pink surface-like structure in the right lung).}
  \label{fig:teaser}
}

\begin{document}


\firstsection{Introduction}\label{sec:introduction}
\maketitle
The Coronavirus disease 2019 (COVID-19) is caused by the SARS-CoV-2 virus and may lead to severe respiratory symptoms (e.g., shortness of breath,  chest pain), hospitalization, ventilator support, and even death.
The real-time reverse transcriptase polymerase chain reaction (RT-PCR) lab test is commonly used for screening patients and is considered the reference standard for diagnosis. However, none of the lab tests, including RT-PCR, are 100\% accurate. For example, sensitivity of RT-PCR depends on the timing of specimen collection~\cite{kucirka2020variation}. Chest CT was found to have significant sensitivity for diagnosing COVID-19~\cite{fang2020sensitivity,islam2020thoracic}.
While the role of chest CT in diagnosis continues to evolve, there is no universal consensus on its usage and recommendation.
Medical practitioners often recommend a chest CT scan as a complement diagnostic test to RT-PCR, or as clinical triage if the patient have severe symptoms that require immediate attention.
For example, chest CT can help in ruling out pulmonary embolism (PE)~\cite{grillet2020acute} which has overlapping symptoms with COVID-19.
Consequently, chest CT remains an important modality for diagnosis, as well as management and prognosis of suspected COVID-19 cases that lead to hospitalization, intensive care unit (ICU) admission, and ventilation. 

Significant amount of research on automatic detection, segmentation, and classification of various diseases (e.g., cancer) in medical imaging have been driven by advances in deep learning 
(DL)
and artificial intelligence (AI).
Similar approaches were also developed recently for detection~\cite{li2020using,panwar2020deep,wang2020weakly,zebin2021covid} and segmentation~\cite{fan2020inf,elharrouss2020encoder,chen2021effective} of COVID-19 lesions on chest X-ray and CT images.
However, these automatic methods remain secondary to the expertise of an experienced radiologist performing manual visual diagnosis on medical images.
Very few applications are currently available that are particularly tailored to support visual diagnosis of COVID-19 that keeps the radiologists in the loop and enhances their workflow.
We have developed \sysname, an elaborate application for visual diagnosis of COVID-19 from chest CT data
that integrates
DL-based automatic analysis with interactive visualization. We incorporate a complete application pipeline from automatic segmentation of lungs, localization of lung abnormalities, traditional 2D views, a suite of contemporary 3D and 2D visualizations suitable for COVID-19 diagnosis,
quantification tools that can help in measuring the volume and extent of the lung abnormalities, and a novel automatic classification model along with visualized activation heatmap that act as a \emph{second reader} of the CT scan.

Our contributions are as follows:
\begin{itemize}
    \vspace{-2mm}
    \item A novel, automatic and effective DL classification model for classifying patients into COVID positive or negative types with good interpretability.
    \vspace{-2mm}
    \item A novel dual visualization-DL application, \sysname, for the diagnosis of COVID-19 using chest CT, above and beyond current workflow of radiologists.
    \vspace{-2mm}
    \item Our \sysname\ incorporates a comprehensive pipeline that includes automatic lung segmentation, lesion localization, novel automatic classification, and a user interface for 2D and 3D visualization tools for the analysis and characterization of lung lesions.
    \vspace{-5.5mm}
    \item \sysname\ incorporates explainable DL and decision support by overlaying the  activation heatmap of our classification model with the 2D views in the user-interface. It also includes quantitative tools for volume and extent measurements of the lesions.
    \vspace{-2mm}
    \item \sysname\ was designed through close collaboration between computer scientists and an expert radiologist who has experience in diagnosing COVID-19 in chest CTs. Important feedback, case-studies, evaluations and discussions with the experienced radiologist regarding our design choices substantially influenced the \sysname\ development and implementation.
\end{itemize}

\section{Related Work}\label{sec:related_work}
A comprehensive visual diagnostic pipeline includes indispensable components
such as the segmentation of relevant anatomy, detection and segmentation of abnormalities,
automatic characterization and analysis of the abnormalities, and interactive visualization tools for qualitative and quantitative analysis.
Each components may use automatic or semi-automatic methods based on desired accuracy, user control, and availability of training data.
Examples of applications include virtual colonoscopy~\cite{hong1997siggraph,hong19953DVC,wang2006improved,franaszek2006hybrid}, virtual bronchoscopy~\cite{haponik1999virtualBroncho,Kiraly2004BronchoPath}, and virtual pancreatography~\cite{jadhav20203dvp,dmitriev2021VA,Dmitriev2017cysts}.
Here, we restrict our discussion to work related to lungs segmentation, visualization and COVID-19 diagnosis.

\subsection{Lungs Segmentation}\label{sec:related:segmentation}
Segmentation is a critical backend operation for visualization and diagnostic applications.
Typically, transfer functions are effective only locally and do not overcome global occlusion and clutter. Segmentation of lungs~\cite{bartz2003hybrid} and its internal anatomical structures, such as the bronchial tree~\cite{kitaoka2002automated, van2008robust} and blood vessels~\cite{zhou2006automatic}, can provide greater control over the 3D rendering and visual inspection tools.
Techniques for vessel segmentation in chest CT range from local geometry-based methods such as vesselness filters~\cite{frangi1998multiscale, jerman2015beyond} that utilize a locally-computed Hessian matrix to determine the probability of vascular geometry, to more sophisticated methods utilizing supervised DL.
Some methods also attempt a hybrid approach to combine vesselness filters and machine learning~\cite{korfiatis2011vessel}. 
Lo et al.~\cite{lo2012extraction} have conducted a comparative study of methods for bronchial tree segmentation.
Other methods focus on segmenting the entire lungs and identifying the interior lobes~\cite{van2010automatic}.
Segmenting healthy lungs can be easier than diseased lungs since the geometry of the lungs and its internal features can change drastically with a disease.

 Segmentation in chest CT plays an essential role in lesion quantification, diagnosis and severity assessment of COVID-19 by delineating regions of interest (ROIs) (e.g., lung, lobes, infected areas). 
 Segmentation related to COVID-19 from chest CT could be categorized into two groups: lung segmentation and lesion segmentation. The popular classic segmentation models including U-Net\cite{ronneberger2015u} and Deeplabv3\cite{chen2017rethinking} and their variants are widely used for COVID-19.  Wang et al.~\cite{wang2020weakly} have trained a U-Net using lung masks generated by an unsupervised learning method and then used the pre-trained U-Net to segment lung regions. Zhang et al.~\cite{zhang2020clinically} have constructed a lung-lesion segmentation framework with five classic segmentation models as the backbone to segment background, lung fields, and five lesions. Fan et al.~\cite{fan2020inf} have developed a lung infection segmentation deep network (Inf-Net) for COVID-19 and proposed a semi-supervised learning framework to alleviate the shortage of labeled data.
 Hofmanninger et al.~\cite{hofmanninger2020automatic} have compared four generic deep learning models (U-Net, ResU-Net, DRN, Deeplab v3+) for lung segmentation using various datasets,
 and have further trained a model for diseased lungs.
 We found this model to be fairly robust for COVID-19 cases, and therefore, we have adapted this model for our \sysname\ application pipeline.

\subsection{COVID-19 Classification}\label{sec:related:classification}

Computer-aided diagnosis (CAD) of COVID-19 can assist the radiologists, as it is not only instantaneous but also can reduce errors caused by radiologists’ visual fatigue and lack of training. The rapid increase in the number of suspected or known COVID-19 patients has posed tremendous challenge to radiologists regarding the increasing amount of work. 
CAD of COVID-19 can act as a second reader and thus help the radiologists.
A common problem for developing a CAD system is weakly annotated data in CT images, where usually only patient-level diagnosis label is available. Some studies developed their diagnosis systems using the result of lesion segmentation ~\cite{zhang2020clinically, shi2021large, xu2020deep}. They firstly trained a lesion segmentation model and then input the segmented lesion to the classification model. However, manual annotation of lesion masks for training the segmentation model is very expensive. Another type of method is slice-based ~\cite{jin2020development, mei2020artificial, bai2020artificial}. The slice-wise decisions obtained by a 2D classification model are fused to get the final classification result for the CT volume. Similarly, the manual selection of the infected slices among all the CT slices is of high cost. Some other diagnosis techniques were developed with 3D convolutional neural networks (CNNs) ~\cite{wang2020weakly, harmon2020artificial}. While 3D CNN can capture the spatial features, the complexity of the 3D convolution makes it harder to interpret. Besides, 3D CNN usually requires more GPU memory, which makes it difficult to be trained on machines with limited GPU memory size.

Here, we build a COVID-19 classification model based on deep multiple instance learning (MIL) to address the problem of weakly annotated data in chest CT. Our \sysname\ integrates this effective and efficient model as a second reader to assist the radiologist. Furthermore, it provides the class activation heatmap generated by Grad-CAM ~\cite{selvaraju2017grad} to visualize important regions used for the classification model decisions, making it more transparent and explainable to the radiologists.

\subsection{Visualization and Diagnostic Systems}\label{sec:related:visualization}
Many visual diagnostic systems for lungs focus on the paradigm of virtual endoscopy~\cite{haponik1999virtualBroncho} and path planning and navigation~\cite{summers1997navigational,Kiraly2004BronchoPath,aguilar2017rrt}.
Region-growing~\cite{bartz2003hybrid,da2017segmentation,cortez20133d} has been proposed for isolating and visualizing the lungs and their internal features.
Lan et al.~\cite{lan2018visually} have used the selection of voxels over intensity-gradient histograms and spatial connectivity for visualizing lungs and their structures.
Automatic semantic labeling of bronchial tree~\cite{kitaoka2002automated,van2008robust} can support further analysis and visualization.
Volume deformation~\cite{santhanam2008modeling} and context preserving planar reformations~\cite{marino2015planar} are deployed for managing occlusions, partial abstraction or visual simplification.
Wang et al.~\cite{wang2019deeporgannet} have proposed DL model for reconstruction of lungs 3D/4D geometry from 2D images (e.g., X-Ray).
Hemminger et al.~\cite{hemminger2005assessment} have developed a 3D lungs visualization application for cardiothoracic surgical planning (e.g., for lung transplant and tumor resection).
To the best of our knowledge, there are no COVID-19 oriented 3D visualization systems that focus on supporting radiologists' visual diagnosis workflow, such as \sysname. Some of the DL-based classification models only incorporate restricted 2D visualizations on CT or X-ray images, in the form of Grad-CAM activation or localization heatmaps~\cite{panwar2020deep,zebin2021covid}.
The Coronavirus-3D visualization system~\cite{sedova2020coronavirus3d} presents only a dashboard for tracking SARS-Cov-2 virus mutations and 3D structure analysis of related proteins.

Many general-purpose open source software (e.g., 3D Slicer~\cite{pieper20043dSlicer, kikinis20143dSlicer}, ParaView~\cite{ahrens2005paraview}, MeVisLab~\cite{heckel2009object}) support image analysis and volume visualization tasks.
These in turn extend their abilities by building upon or integrating open source libraries (e.g., VTK~\cite{schroeder2006VTK,schroeder2000VTKTutorial}, ITK~\cite{ibanez2003itk}) that provide an extensive breadth of image and geometry processing, and rendering techniques.
Our \sysname\ is specifically designed with chest CT inspection in mind.
We implement \sysname\ ground-up using VTK and Qt with a simplified and essential interface. It could have also been implemented using other open source frameworks (e.g., ParaView, 3D Slicer, or MeVisLab).
However, beyond implementing a specialized application, our contributions in \sysname\ are the selection and curation of essential tools and interface for chest CT inspection through collaboration with expert radiologists who are experienced in inspecting chest CT for COVID-19. Our pipeline also seamlessly integrates lungs and lesions automatic segmentation, novel COVID-19 classification, and incorporates appropriate results from these models (classification probabilities, activation maps, automatic measurements) that do not come out-of-the-box in general software frameworks.

\section{COVID-19 Background and Task Analysis}\label{sec:background}
While there is no universal consensus on the usage of chest CT for COVID-19 diagnosis,
some practitioners request chest CT for patient management, both confirmed and unconfirmed cases, to complement lab tests (RT-PCR) which suffer from inaccuracies. To improve the accuracy of CT, various approaches have been reported. For instance, Fan et al.~\cite{pan2020time} divided the course of COVID into four temporal stages to account for differing CT findings.
A recent report~\cite{ruch2020ct} shows that lung involvement (lesion volume percentage with respect to the lungs) is a good predictor of patient outcome in terms of ICU admission and death.
In this section, we discuss the manifestation and appearance of COVID-19 lung abnormalities in chest CT as relevant to visual diagnosis.
However, these imaging features are not unique to COVID-19 (not pathognomonic) and can be caused by other infections.

Particularly, our system design choices focus on four prominent lung lesions:
ground glass opacities (GGOs), consolidations, interlobular septal thickening (IST), and vascular pathology (i.e., dilation of blood and air vessels).
There are numerous other reported abnormalities/lesions~\cite{Homsi2020review,kwee2020chest}, but they occur less frequently in COVID-19.
Apart from the lesion type, its locations, distribution, and left-right lung symmetry are also critical in assessing the patients.

\subsection{COVID-19 Chest-CT Imaging Features}\label{sec:ct_features}
\begin{figure}
    \centering
    \includegraphics[width=0.9\linewidth]{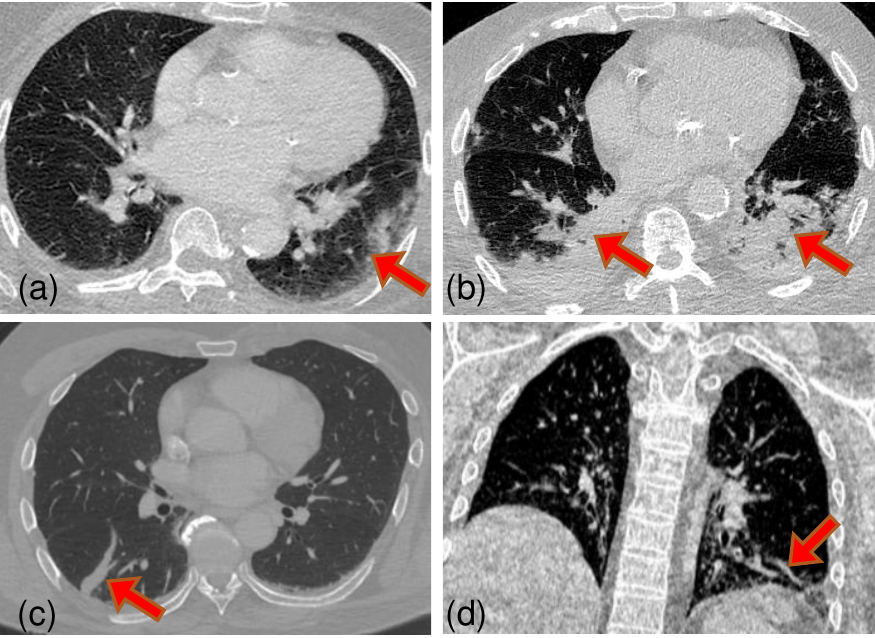}
    \caption{Prominent chest CT features for COVID-19: (a) GGO. (b) Consolidations. (c) Interlobular septal thickening. (d) Vessel enlargement.}
    \label{fig:ct_features}
\end{figure}

\textbf{Ground Glass Opacity (GGO):}
GGO is the most common chest CT feature of COVID-19.
It appears as low intensity regions (as compared to lung vessels) around the lung vessels usually in the periphery of the lungs. Fig.~\ref{fig:ct_features}a 
shows an example axial image with GGO.
Early stage GGO is often unilateral (i.e., in one lung), whereas intermediate and late stage GGOs often have peripheral and bilateral distribution.

\textbf{Consolidations:}
Consolidations are formed when GGOs translate into denser regions over time. They appear as solid mass on the CT images (see
Fig.~\ref{fig:ct_features}b).
Similar to GGOs, they have peripheral distribution, unilateral in early stage and bilateral in intermediate and late stage.

\textbf{Interlobular Septal Thickening (IST):}
The septal walls between lobes can thicken due to COVID-19. They are difficult to locate, as they are very similar to blood vessels in the planar views (see Fig.~\ref{fig:ct_features}c).

\textbf{Vascular Pathology:}
Blood vessels within the lungs can show abnormalities such as enlargement or dilation (see Fig.~\ref{fig:ct_features}d), 
particularly within the neighborhood of other abnormalities (e.g., GGO). Similarly, the bronchial tree vessels (air-ways) show abnormally thicker walls.
These imaging features are very subtle and not always discernible.

Additional imaging features related to COVID-19 include intralobular septal thickening which often appears as \emph{crazy paving pattern} on the planar images, and mixing of GGO and consolidation morphology resulting in halo and reverse-halo signs. A comprehensive list of features and their frequency of occurrences in COVID positive patients is discussed by Homsi et al.~\cite{Homsi2020review} and by Kwee et al.~\cite{kwee2020chest}.

\begin{figure*}[ht]
    \centering
    \includegraphics[width=0.9\linewidth]{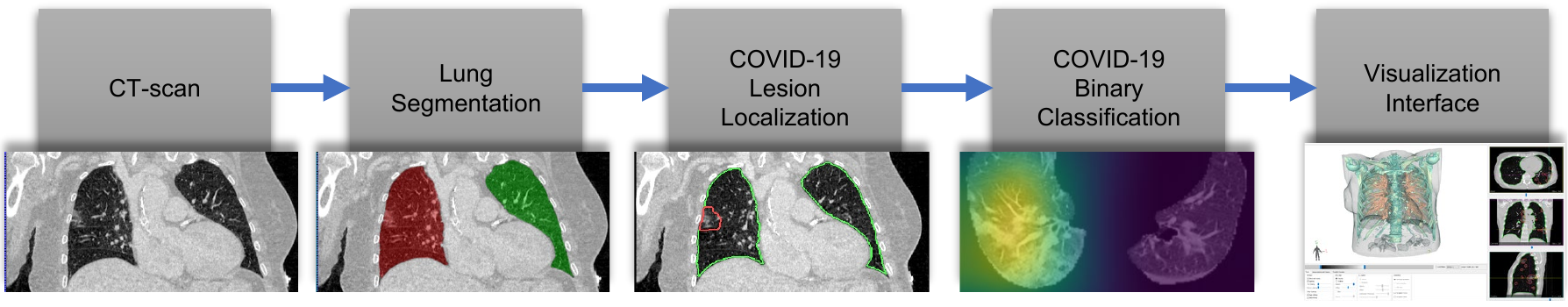}
    \caption{The \sysname\ application pipeline. 
    The segmented lungs and lesion localization from the CT scan are processed by our COVID-19 classifier, which calculates the probability of COVID-19 positive/negative for the case, and generates an attention heatmap, as part of the explainable DL, that is displayed along with other exploratory 3D/2D visualizations in the user interface.} \label{fig:pipeline}
\end{figure*}

\subsection{Requirements and Task Analysis}\label{sec:task_analysis}
We discuss here the conventional workflow of radiologists while diagnosing COVID-19 on chest CT scans, and determine high-level tasks that are commonly performed.
The analysis of 2D chest CT images follows a largely conventional workflow, though the tasks performed and their order may vary between practicing radiologists.
A chest CT scan is ordered not only for the singular task of diagnosing COVID-19 and its severity, but also due to other possible conditions that may require urgent attention, such as pulmonary embolism (PE).
Thus, apart from examining the lungs, the radiologist often performs a holistic analysis of other regions, such as the heart and vascular structures in the lungs proximity, swelling of axillary lymph nodes, as well as searching for the presence of other lesions and abnormalities throughout the CT scan.

In the conventional 2D workflow, the radiologist may begin with identifying the range of slices (and extents within the slices) within which the lungs are contained, by observing the extremities and lungs boundary (Task T1). This task also includes inspection of the lung boundaries for abnormal/stiffness of shape.
The radiologist will also adjust the gray scale map to improve contrast between the lungs background and the vascular structures such as the bronchial tree (Task T2), the arteries, and the veins inside the lungs.

The radiologist then scans through the identified range of 2D axial planes for lung abnormalities listed in Sec. \ref{sec:ct_features},
the most common of which are GGOs and consolidations (Task T3). In addition, the radiologist will determine the location and distribution of these abnormalities (Task T4). Peripheral and bilateral distributions are characteristic of COVID-19 diagnosis.
The radiologist will also look for more subtle abnormalities, such as interlobular and intralobular septal thickening and vascular enlargements (Task T5).
Early stage abnormalities can be subtle and hard to detect, which requires careful and comprehensive inspection of the lungs.
Furthermore, the radiologist may measure the lesions (e.g., GGO) to determine the growth or severity of the disease (Task T6), helping in patient management and prognosis.
A radiologist experienced in COVID-19 diagnosis confirmed these high-level tasks, and that currently they do not use any 3D visualization tools.

While these tasks are identified in the conventional 2D workflow, our design choices and visualization tools translate these tasks to include them in the combined 3D/2D workflow.
3D visualization of the segmented lungs can provide a holistic view of lungs and lesions to identify the distribution of affected areas. While a similar assessment can also be made using 2D slice views, our 3D visualizations provide an alternate viewpoint that can further inform the radiologist beyond their usual workflow. As shown in our case studies~\hyperref[paragraph-case3]{(Case 3)}, 3D can also make IST identification easier than in 2D views. Additionally, it has been generally accepted that 3D visualizations can be better in inspecting vascular structures such as bronchial trees and blood vessels in the lungs and elsewhere. While GGO and consolidations are macro-level structures easily seen in 2D slices, inspection of more subtle abnormalities such as smaller opacity regions, vascular enlargements, and septal thickening can benefit from 3D visualizations.

In addition to supporting these identified tasks, we design our system to fully support their conventional 2D workflow.
This also helps in rare events when the system fails to compute the necessary information such as the segmentation masks. In such cases, the radiologist can still continue to analyze the case using their conventional 2D workflow.

\section{The \sysname\ Application}\label{sec:application}
We have designed the \sysname\ application based on the requirements of the COVID-19 chest CT diagnosis and task analysis performed in
Sec. \ref{sec:background}.
The high-level pipeline of \sysname\ is shown in Fig. \ref{fig:pipeline},
and incorporates all the important components for such a visual diagnostic system, including automatic lung segmentation, automatic lesions and abnormalities segmentation, and novel ML model for classification of cases into COVID-19 negative/positive.
All the modules are integrated into a single application and presented through a well-designed user interface. Since the modules use fully-automatic methods, the user does not have to interact with any of the pre-processing steps before the dataset and the processed information is loaded into the visualization interface.
The lungs segmentation mask, lesion segmentation, and classification results are computed only once when a new dataset is loaded into the system. These results are then cached into the hard drive so that future loading and analysis of the CT scans happens faster,
as re-computing these for every execution is time-consuming and unnecessary.
We discuss each of the \sysname\ modules and its user-interface and visualization tools in the following subsections.

\subsection{Lungs Segmentation}\label{sec:segmentation}
Use of a segmentation mask in a 3D rendering pipeline can provide significant control over the visualization process through deployment of local transfer functions and multi-label rendering.
Global transfer functions have limited ability to control rendering occlusions and clutter in complex datasets (e.g., chest and abdominal CT scans).
Existing visual diagnostic systems, such as virtual colonoscopy~\cite{hong19953DVC} and virtual pancreatography~\cite{jadhav20203dvp} also incorporate organ segmentation methods.

In \sysname, we incorporate Hofmanninger et al.~\cite{hofmanninger2020automatic} lungs segmentation model that was trained on a large variety of diseased lungs, encompassing different lesions and abnormalities with air pockets, tumors, and effusions, and includes COVID-19 data. In our tests, we found this model to be fairly robust on COVID-19 datasets.
Particularly, we did not find any cases where the segmentation outline deviated significantly from the true boundary of the lungs. Any rare and minor deviations did not degrade the utility of our analysis pipeline.
This incorporated automatic segmentation allows us to provide segmentation outlines in 2D views, compute the lungs clipping box, and 3D render the lungs anatomy in isolation, which supports Task T1 (see Sec. \ref{sec:task_analysis}).
Hofmanninger et al. also provide a model for lungs lobe segmentation, which is desirable for our system as it can support identification of anatomical locations for the COVID-19 lesions. However, the model was not found to be very robust on lungs with large COVID-19 lesions.
In case of significant GGO and consolidation, the model failed to provide
satisfactory segmentation of the lobes.
Inaccurate segmentation in a diagnostic application can lead to degraded 3D visualizations and incorrect diagnosis. 
Thus, we prefer models that work more reliably even though they may not provide further subdivision of structures.

\subsection{Lesion Localization and Detection}\label{sec:localization}
Our system provides segmentation outlines in 2D views for identifying and examining regions of abnormalities, drawing the radiologist's attention to regions that should be examined more closely.
Localization or segmentation of COVID-19 lesions allows us to provide these outlines in 2D views.
The segmentation mask is also used to render the lesions in 3D view and to highlight them along with the lungs surrounding structures.
We have adapted and integrated into our pipeline, a COVID-19 lesion segmentation model by Fan et al.~\cite{fan2020inf}.
They provide a binary segmentation model that segments overall lesions and abnormalities of the lungs as well as a multi-class model that is trained to segment both GGO and consolidation regions. We found that on our datasets, this model worked more accurately to highlight the regions of abnormality, which is consistent with the dice overlap numbers (0.739 and 0.458 for binary and multi-class models, respectively) reported by Fan et al.
The binary detection model that we utilize (Semi-Inf-Net) is trained to identify lung abnormalities using COVID-19 chest CT images that predominantly contain GGOs and consolidations, and is thus suitable for our needs.
After localization, the characterization of the lesions into further sub-types is handled by the radiologist through examination using the visualization tools provided in the user interface.

\subsection{Classification}\label{sec:classification}
Multiple instance learning (MIL) is a kind of weakly supervised learning~\cite{zhou2018brief}. It was first formulated by Dietterich et al.~\cite{dietterich1997solving} for drag activity prediction and then widely applied to various tasks. In MIL, the training set consists of bags, where each bag is composed of a set of instances. The goal is to train a model to predict the labels of unseen bags. In MIL, only the bag-level label is given, and the instance-level label is unknown. This setting is particularly suitable for medical imaging, where typically only image-level or patient-level label is given. According to Amores' taxonomy~\cite{amores2013multiple}, MIL algorithms can be categorized into three groups: Instance-Space (IS) paradigm, Bag-Space (BS) paradigm, and Embedded-Space (ES) paradigms. The IS paradigm learns an instance-level classifier and the bag-level classifier is obtained by aggregating the instance-level response. The BS and the ES paradigms treat each bag as a whole entity and learn a bag-level classifier by exploiting global, bag-level information. The difference between both paradigms is how the bag-level information is extracted. The BS paradigm implicitly calculates the bag-to-bag similarity by defining a distance or kernel function, while the ES paradigm explicitly embeds the bag into a compact feature vector by defining a mapping function.

MIL pooling methods are used to represent the bag with the corresponding instances. Popular methods include max, mean, and log-sum-exp poolings~\cite{wang2018revisiting}. However, they are non-trainable which may limit their applicability. To address this, some methods such as noisy-and pooling ~\cite{kraus2016classifying} and adaptive pooling ~\cite{zhou2017adaptive} have been developed, but their flexibility is restricted. Ilse et al.~\cite{ilse2018attention} have proposed an attention-based pooling that is fully trainable and can weigh each instance for the final bag prediction. Han et al. \cite{han2020accurate} applied this pooling\cite{ilse2018attention} for bag representation, and a 3D CNNs was designed by Wang et al.\cite{wang2020weakly} as deep feature generator in their COVID-19 CAD system. Here, we also build our COVID-19 classification model based on this pooling method ~\cite{ilse2018attention} for its flexibility and interpretability. In addition, we propose a regularization term composed of differences between learned weights of adjacent slices to further smooth the attention weights and enhance the connection between adjacent slices.
Also, we use ResNet18\cite{he2016deep} to transform each slice into feature vector for its strong feature extraction ability, and thus
each slice contains more information for diagnosis. Our slice-based deep MIL method can thus generate the class activation map with Grad-CAM\cite{selvaraju2017grad} for each slice to increase our model interpretability and decision support. Our classification model performance is also superior due to the fact that each slice contains more information. See Sec. \ref{sec:eval:classification} for quantitative results of our classification model.

In our COVID-19 DL classification, we denote each CT scan as a bag-label pair $\{X,Y\}$, where $X=\left\{\mathbf{x}_{1}, \ldots, \mathbf{x}_{K}\right\}$ denotes the CT volume containing $K$ instances, instance $\mathbf{x}_{k}$ denotes the slice, and $Y \in\{0,1\}$ denotes whether the patient is COVID-19 positive.
$K$ can vary for different bags. We use a feature extractor parameterized by the CNNs $f_{\psi}(\cdot)$ with parameters $\psi$ to transform the instance $\mathbf{x}_{k}$ into a low dimensional feature vector $\mathbf{h}_{k}=f_{\psi}\left(\mathbf{x}_{k}\right)$ where $\mathbf{h}_{k} \in \mathbb{R}^{D}$. 
Here, we use the ResNet18 as $f_{\psi}(\cdot)$ and $D=512$. For the bag presentation, we use the following attention-based pooling method ~\cite{ilse2018attention} as the mapping function in the Embedded-Space paradigm:\begin{equation} \label{eq:1}\mathbf{z}=\sum_{k=1}^{K} a_{k} \mathbf{h}_{k},\end{equation}where:\begin{equation} \label{eq:2}a_{k}=\frac{\exp \left\{\mathbf{w}^{\top} \tanh \left(\mathbf{V} \mathbf{h}_{k}^{\top}\right)\right\}}{\sum_{j=1}^{K} \exp \left\{\mathbf{w}^{\top} \tanh \left(\mathbf{V} \mathbf{h}_{j}^{\top}\right)\right\}}\end{equation}where $\mathbf{w} \in \mathbb{R}^{L \times 1}$ and $\mathbf{V} \in \mathbb{R}^{L \times D}$ are parameters of a two-layered neural network. Here we set the dimension $(L)$ in $\mathbf{V}$ as 128. Let $L_{CE}$ denote the binary cross entropy loss as follows:  \begin{equation} \label{eq:3}L_{\mathrm{CE}}=-\frac{1}{N} \sum_{i=1}^{N} \sum_{c=0}^{1} p\left(Y_{i}=c\right) \log \left(q\left(Y_{i}=c\right)\right),\end{equation} where $N$ is the batch size, $p\left(Y_{i}=c\right)\in\{0,1\}$ is the true class probability of $X_{i}$ belonging to the class $c$ and $q\left(Y_{i}=c\right)$ is the estimated class probability of $X_{i}$ belonging to the class $c$ . 
The loss function is:
\begin{equation} \label{eq:4}L_{Total}=L_{CE}+\lambda L_{AW},\end{equation}where: \begin{equation} \label{eq:5}{L}_{AW}=\sum_{i=2}^{K}\left(a_{i}-a_{i-1}\right)^{2}\end{equation} and $\lambda$ is a non-negative constant to balance $L_{CE}$ and $L_{AW}$. Eqs. \ref{eq:4} and \ref{eq:5} are inspired by the similarity between adjacent slices. As two adjacent CT slices are similar, their learned instance weights should also be similar. Thus, the difference between two adjacent slices weights should be very small. The proposed regularization term $L_{AW}$ facilitates the attention-based pooling module assigns the weights to each instance better, and thus can further improve the model performance on bag prediction.

\begin{figure}[ht]
    \centering
    \includegraphics[width=\linewidth,trim={0 60 0 60},clip]{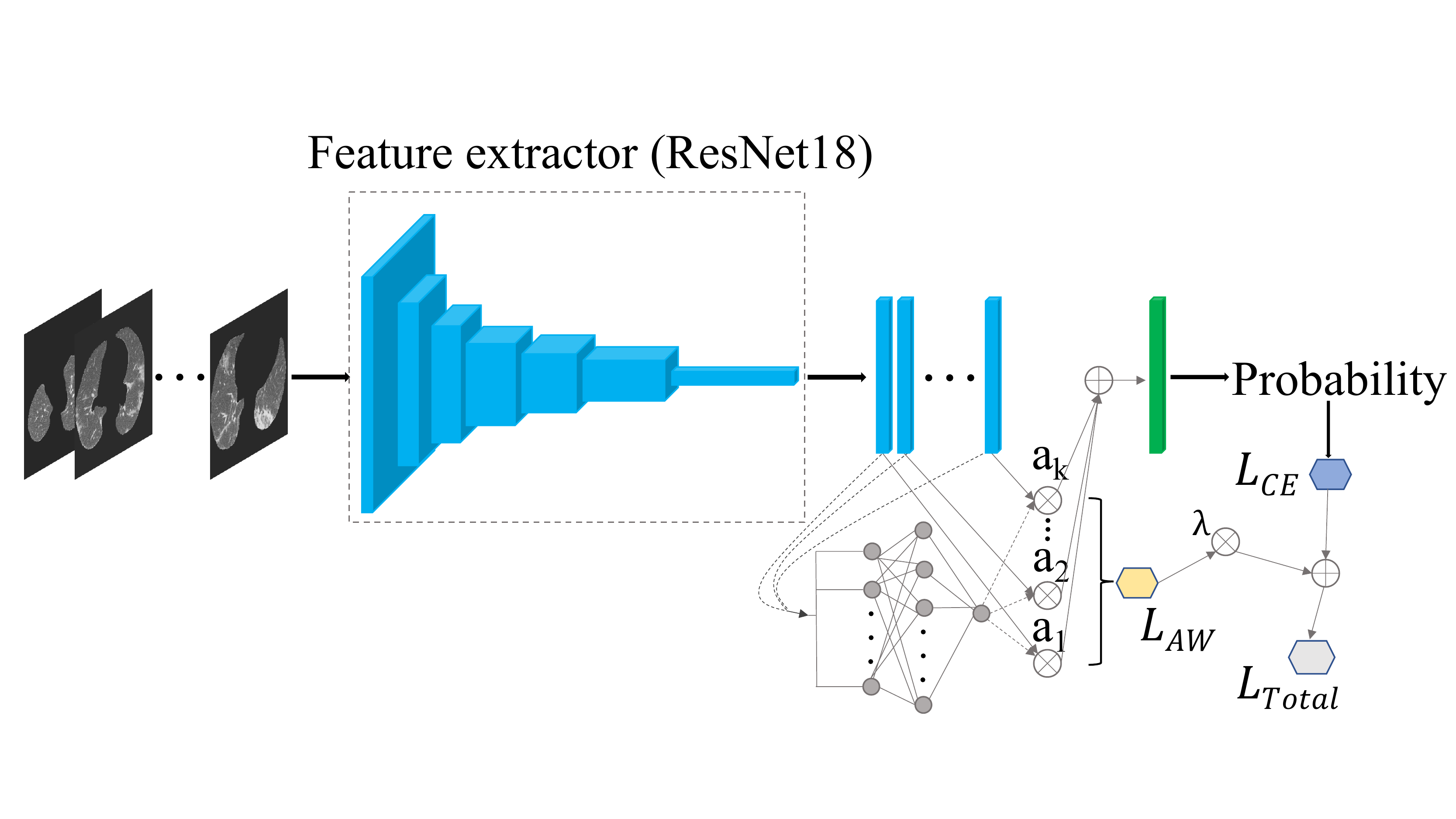}
    \caption{Architecture of our COVID-19 classification model.}
    \label{fig:model_diagram}
\end{figure}

As shown in Fig. \ref{fig:model_diagram}, the classification model takes the preprocessed CT volume as input (details later), and the feature extractor transforms each slice into a low dimensional feature vector. All the feature vectors are mapped into a semantic embedding representing the CT information using the attention-based pooling method mentioned above. Then, the embedding is further processed by a {fully connected (FC) layer and softmax function} to output the probability of different classes (COVID-19, non-COVID-19) for the CT volume. Eqs. \ref{eq:4} is used to calculate the total loss $L_{Total}$ and the model is trained end-to-end by backpropagation.

Data prepossessing starts with 
the CT images extracted from the DICOM files, and the CT volume is resampled to the same spacing of 1mm in the $z$-direction. We used the lung segmentation mask generated by a pre-trained U-Net \cite{hofmanninger2020automatic} to extract the lungs and remove the background. Then, a bounding box is calculated using the lung mask to crop the lung region. The bounding box is padded to keep the width and height of all CT images the same. The original CT intensity values are clipped into $[- 1250, 250]$ range and then normalized into $[0, 1]$. Next, the CT images are resized to $T \times 224 \times 224$, where $T$ is the number of slices. To reduce overfitting, online data augmentation strategies, including random rotation ($-10$ to $10$ degrees) and horizontal/vertical flipping with $50\%$ probabilities are applied. For each training example, the augmentation is the same for every slice in the volume.

We qualitatively compare our classification model with other existing COVID-19 detection algorithms. Unlike~\cite{zhang2020clinically, shi2021large, xu2020deep, jin2020development, mei2020artificial, bai2020artificial} requiring lesion segmentation or manual selection of infected slices, our method only requires patient level weak label, which is much easier to obtain. Li et al. \cite{li2020artificial} have proposed 2D CNNs to extract features of each slice, then the slice-wise features were fused into CT volume-level feature via a max-pooling layer. However, since max pooling is non-trainable, its applicability is limited. Ilse et al. \cite{ilse2018attention} have shown that attention-based pooling outperforms max pooling in image classification experiments. Besides, the attention weights for each slice makes the model more interpretable. Thus, attention-based pooling used in our model is more effective and interpretable than max pooling. Furthermore, our proposed regularization term helps the attention-based MIL pooling module to assign the weights to each instance better by considering the similarity between adjacent slices, thereby can further improve the model classification performance. There are also 3D CNNs methods~\cite{wang2020weakly, harmon2020artificial}, but, 3D CNNs usually require larger GPU memory. Also, our framework can be used with widely available 2D CNNs pretrianed on ImageNet \cite{deng2009imagenet}, which makes the convergence of model training faster and better. Furthermore, the quantitative weights learned for each instance and Grad-CAM make our model more interpretable, which is helpful for diagnosing and improving the model.
\begin{figure*}[ht]
    \centering
    \includegraphics[width=\linewidth]{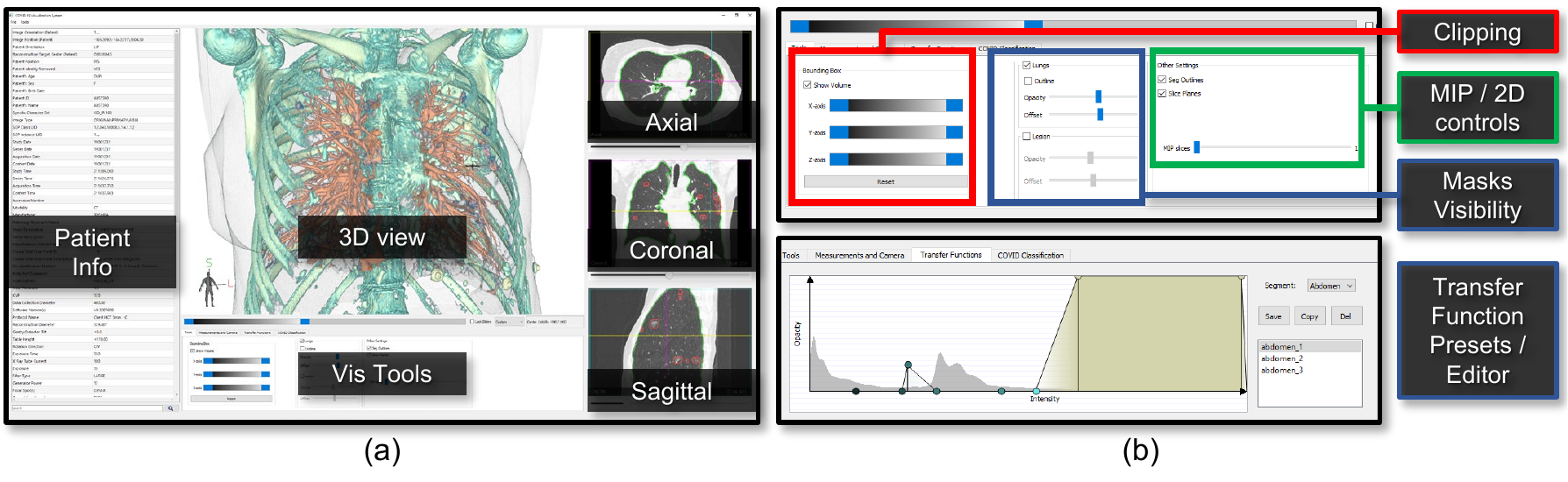}
    \vspace{-6mm}
    \caption{Snapshot of \sysname\ user interface. (a) The top panel of the user interface contains the central 3D view, visualization tools, and the conventional 2D axial, sagittal and coronal views. Vertical panel on the left shows DICOM patient information. (b) The bottom panel of the user interface provides widgets for controlling the 2D/3D visualizations, clipping, measurements, and access to classification model results and heatmap.}
    \label{fig:ui}
\end{figure*}

\subsection{Visualization Tools and User Interface}\label{sec:visualization}
The visualization interface of \sysname\ integrates all the components of our pipeline (Fig. \ref{fig:pipeline}) for the radiologist to access in different modes.
The interface allows the radiologist to visualize the chest CT in different 2D and 3D views that support conventional 2D radiologists' workflow and contemporary 3D visualization methods for better discernment of lesions and their morphology for improved diagnosis.

\textbf{2D Views:}
A snapshot of the user interface of \sysname\ is shown in Fig. \ref{fig:ui}a.
The right-hand side displays the conventional 2D views of axial, coronal, and sagittal planes.
A range-slider immediately below the \emph{3D View} is used for adjusting the gray-scale colormap
in all the 2D views. Similar action can also be performed using mouse left-click and drag operation on the 2D views. This supports Task T2.
These views are indispensable as radiologists are trained to use them to examine scans.
We incorporate additional cues in the 2D views to highlight the segmented lungs outline (in green) and COVID-19 lesions (in red),
which support Tasks T1 and T3, respectively.
Thus, we support the radiologists in their conventional workflow (Tasks T1 and T2) and augment additional information to draw attention to abnormal regions of the lungs that require special attention and characterization.

\begin{figure}[ht]
    \centering
    \includegraphics[width=0.8\linewidth]{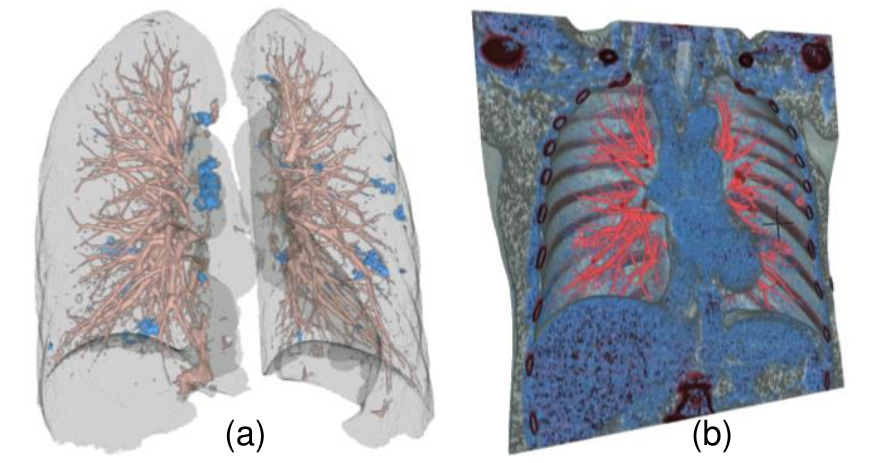}
    \caption{Lungs multi-label 3D visualization. (a) Rendering lungs and lesions as volume and lungs outline as surface. (b) Rendering lungs with outer context volume and coronal clipping plane for managing occlusion. 
    The heart chambers and  subcutaneous fat can be seen with the clipping.
    }
    \label{fig:3d_eg}
\end{figure}

\textbf{3D View:}
In Fig. \ref{fig:ui}a, the main canvas shows the 3D rendering of the chest CT, 
utilizing multi-label volume rendering and applies local transfer function to each segmentation label.
The composite segmentation mask used by the 3D rendering contains the segmented lesions, segmented lungs, and the context volume outer region.
Each of the three regions use three separate transfer functions.
Additionally, the user can choose to hide/show each of these regions in the 3D view, thereby
allowing to create different visualization combinations for better understanding of the lungs and surrounding regions (Task T3).
For example, the user may choose to hide the context volume and only render the lungs and internal lesions to get an occlusion-free view of the lungs interior (Fig. \ref{fig:3d_eg}a).
As seen in the figure, we also allow the user to render the lungs outline geometry as a translucent surface mesh. This provides important context when rendering partial or restricted volumes in the 3D view and also supports Task T1 for inspection of lung geometry for stiffness or restrictions.
As another example, the user may choose to render all three regions and use one or more clipping planes to control occlusion and get a look into the lungs/chest (Fig. \ref{fig:3d_eg}b).
The 2D views and the 3D view are linked to each other for easier navigation, correlating features, and simultaneous lesion inspection across views. Clicking on any one of the planes will steer the other two planes to the clicked voxel, and a 3D cursor crosshair (3 black orthogonal intersecting lines) will update its position in the 3D view to the selected voxel position.
Similarly, the user can directly select a point in the 3D view by clicking twice from different viewpoints while pressing the control key. This will update the 3D cursor, 
and the 2D view points will automatically steer to the corresponding
axial, coronal, and sagittal planes that intersect with the selected voxel.

\begin{figure}[ht]
    \centering
    \includegraphics[width=0.9\linewidth]{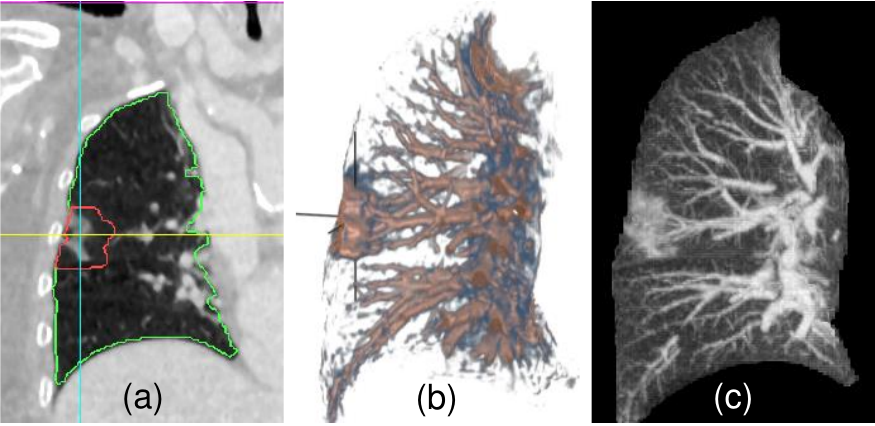}
    \caption{Thick slab mode can provide 3D extensions of the conventional 2D planar views. (a) Coronal view of slice 232. (b) Thick slab mode in coronal view. 3D cursor cross-hair points to a GGO whose morphology is clearly visible in 3D rendering along with neighboring vessels that connect with it. (c) MIP view in coronal plane using 10 adjacent slices around slice number 232. GGO lesions have a much larger footprint in MIP view and hence are easier to spot. The single slice views (a) only show fragments of the lesion and is difficult to judge the shape and morphology of the lesion in conventional 2D views.
    (a, c) Comparison of MIP mode with conventional 2D views for lesion visualization.}
    \label{fig:thick_slab}
\end{figure}

\textbf{Clipping Tool:}
As shown in Fig. \ref{fig:3d_eg}b, we include a volume clipping tool that works in tandem with the multi-label volume renderer, which can render three different regions (context, lungs, lesions) with localized transfer functions.
The clipping tool uses three range sliders, one for each axis, to control the six clipping planes (also covering Task T1).
The user can choose to move any single clipping plane at any given time (e.g., Fig. \ref{fig:3d_eg}b uses single coronal clipping plane), or both planes of an axis by dragging the middle of the range sliders.
This function of dragging both minimum and maximum clipping plane of an axis allows a visualization mode
where a thick slab of the volume is rendered and the slab can be moved along the axis for 3D inspection of the lung slabs. This can be considered as the 3D extensions of the axial, sagittal and coronal 2D views.
The thick slab mode allows better discernment of the local 3D structures and lesion morphology without significant surrounding occlusion (Fig. \ref{fig:thick_slab}). This can support task T5 for closer inspection of lesion morphology and understanding subtle geometries.

\textbf{Transfer Function Design and Presets:}
\sysname\ provides two different ways to manipulate optical properties of 3D rendering: basic mode, and advanced mode. In the basic mode, the user manipulates two sliders: opacity and offset.
The opacity slider modifies the global opacity of a label (lungs, lesions, or context volume).
Similarly, the offset slider applies an offset to the local transfer function of the chosen label.
Specifically, the mapping between the transfer function and the scalar range over which it is applied can be manipulated using this slider. It \emph{offsets} the transfer function mapping to lower or higher values of intensity.
This allows for easy manipulation and adjustment of the preset transfer functions without the need
to directly manipulate the piece-wise linear color and opacity maps.
In the advanced mode, the user can choose the \emph{Transfer Function Tab} in the \emph{Vis Tools}  (Fig. \ref{fig:ui}a)
to directly edit the transfer functions as a polyline on a 2D graph of Intensity vs Opacity (Fig. \ref{fig:ui}b).
We provide some well-designed preset transfer functions that the user can directly choose for
visualizing the context volume, lungs, and the lesion volume.
The user can also create their own transfer functions and save them for future use.
All saved presets are automatically loaded into the \sysname\ system during future runs. Task T2 of manipulating the colormap in the conventional workflow can be translated to the 3D view as manipulation and management of optical properties or transfer functions. Therefore, this feature of our user interface design accommodates abstract Task T2. 

\textbf{MIP Mode:}
Maximum intensity projection (MIP) mode creates 2D projection of volumes by projecting the highest intensity voxels to the foreground. This rendering is particularly useful for visualizing vascular structures, and consequently are suitable for lungs visualization.
The MIP mode can be activated in all three 2D views, and is applied within the segmented lungs volume.
This helps in overcoming any occlusion caused by the context volume, and the radiologist can focus only on the features internal to the lungs.
Fig. \ref{fig:thick_slab}c-d shows a comparison of MIP mode with conventional 2D views for the visualization of COVID-19 chest CT lesions. 
The MIP mode also essentially supports Tasks T3 and T5 of observing both vascular and GGO/consolidation abnormalities.

\begin{figure}[ht]
    \centering
    \includegraphics[width=0.9\linewidth]{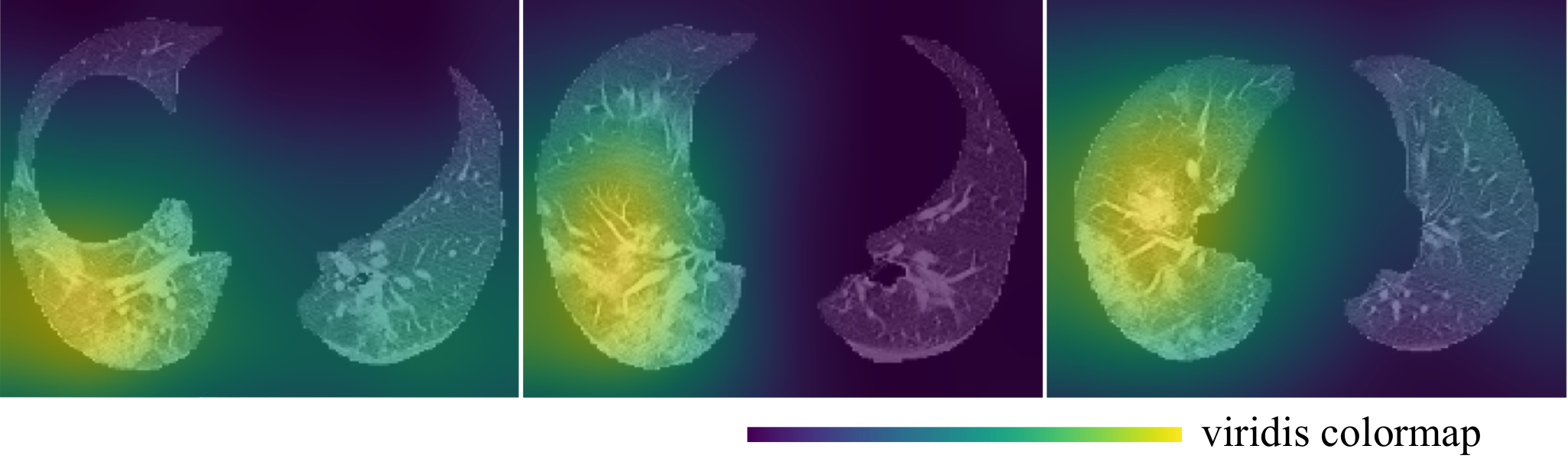}
   \vspace{-1mm}
    \caption{Representative examples of the class activation heatmap for the CT images of COVID-19 patients. }
    \label{fig:heatmap_example}
\end{figure}

\textbf{Explainable DL:}
As automatic classification and assessment models are developed and incorporated into medical diagnosis workflow, it has become important to provide reliable and explainable results to the users.
We described our novel binary classification model for COVID-19 in Sec. \ref{sec:classification}.
The model executes automatically when a chest CT is loaded into the \sysname\.
The results are presented in the form of two percentage probabilities for the COVID negative and positive classes, within a separate \emph{Classification} tab of the \emph{Vis Tools}. We also provide the user with a checkbox to overlay the activation heatmap of our classification model using an adaptation of the Gradient-weighted Class Activation Mapping (Grad-CAM) \cite{selvaraju2017grad} on the 2D axial, coronal, and sagittal views.
We incorporated this activation heatmap as a visual overlay in our user-interface to improve the radiologists' trust in the output of our classification model.
The activation heatmap is extracted from our classification model without any external supervision and allows the radiologist to evaluate what regions of the CT images triggered the classifier results. This provides an explanation and insight into our model results . Fig. \ref{fig:heatmap_example} shows several examples of the activation heatmap in our application for CT images of COVID-19 patients.

\textbf{Measurements:}
\sysname\ provides quantification tools for measuring lesions and tracking their growth.
Linear measurements are supported in all the 2D views. The user can select the \emph{Measurements and Camera} tab in the \emph{Vis Tools} to show, hide, clear, or start linear measurements.
In addition, since we have a lesion segmentation model integrated into \sysname\, we
also provide automatic volume measurement of the lungs and lesions.
Three values are presented: lungs volume, lesions volume, and lesions percentage.
Such volume measurements can help the radiologist to assess the lesion severity and distribution.
As described earlier in Sec. \ref{sec:background}, a recent study~\cite{ruch2020ct} has shown that disease severity based on approximate lesion volume percentage is a promising predictor for patient management and prognosis.
Together, the linear and volume measurements support Task T6.

\section{Evaluation}\label{sec:evaluation}
In this section, we present a quantitative evaluation of our novel COVID-19 classification model,
and a qualitative evaluation of our visualization system and user interface through expert feedback and case studies performed using our system by collaborating radiologists.
\subsection{Classification}\label{sec:eval:classification}
\begin{figure}[ht]
    \centering
    \includegraphics[width=\linewidth]{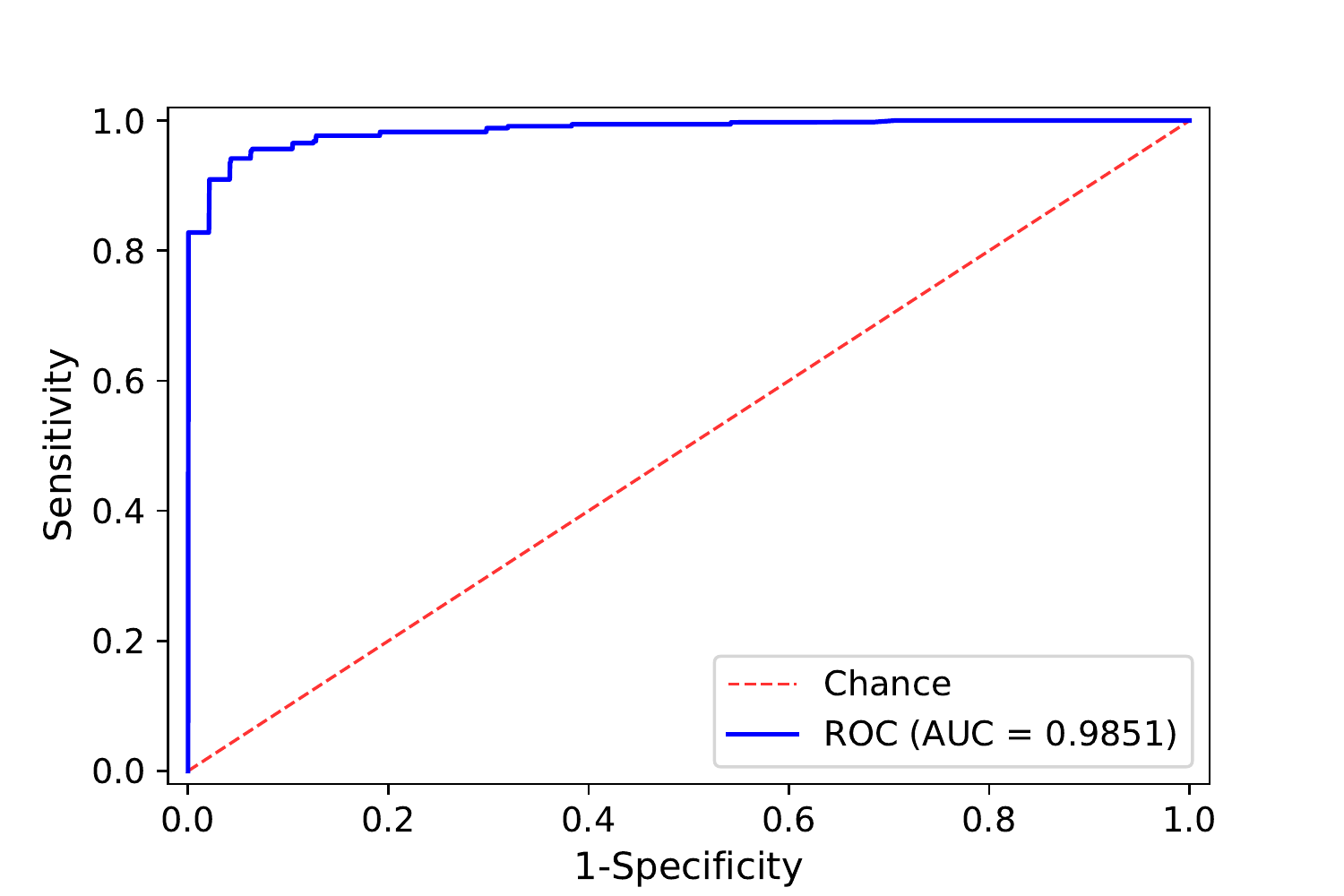}
    \caption{ROC curve of our COVID-19 classification results.}
    \label{fig:ROC}
\end{figure}

Our classification model was developed, trained and evaluated on a total of 580 CT volumes including 343 COVID-19 positive volumes and 237 COVID-19 negative volumes. The dataset size of our study is similar to other studies \cite{wang2020weakly, han2020accurate}. Since public datasets typically have some limitations, such as only positive cases are available or the labels are not RT-PCR-based  \cite{morozov2020mosmeddata}, all our CT scans were collected in our own hospital, as other studies did (e.g., \cite{wang2020weakly, han2020accurate, li2020artificial}).
Our 343 COVID-positive volumes were from patients with RT-PCR positive confirmation for the presence of SARS-CoV-2, and 202 COVID-negative volumes were collected from trauma patients undergone CT exams before the outbreak of COVID-19, and 35 COVID-negative volumes were collected in 2020 from patients without pneumonia. The model was implemented with the PyTorch~\cite{paszke2019pytorch} framework. It was trained using Adam~\cite{kingma2014adam} optimizer with the initial learning rate of $1e-5$ for 100 epochs. We evaluated the performance of the model using 5-fold cross-validation. For the detection of COVID-19, the accuracy and area under the receiver operating characteristic (ROC) curve (AUC) are 0.952( 95\% confidence interval (CI): 0.938, 0.966) and 0.985 (95\% CI: 0.981, 0.989), respectively. The sensitivity and specificity are 0.953 (95\% CI: 0.932, 0.974) and 0.949 (95\% CI: 0.928, 0.97), respectively. The ROC curve of the COVID-19 binary classification results was shown in Fig. \ref{fig:ROC}.
\vspace{-2mm}

\subsection{Expert Feedback and Case Studies}\label{sec:feedback}
We developed  \sysname\  through close collaboration between computer scientists and a co-author expert radiologist (MZ). MZ provided feedback on our design choices during multiple discussion sessions through different development stages.
We gathered a final round of feedback on the design and utility of different tools of our completed system from MZ, another radiologist Dr. Almas Abbasi (AA), and a medical trainee Joshua Zhu (JZ).
Both MZ and AA used \sysname\ on real-world patient cases over remote meetings before providing qualitative feedback.
MZ also performed case studies on multiple real-world cases using the completed {\sysname}.
Following is a description of the case studies and their diagnostic findings.

\begin{figure*}[ht]
    \centering
    \includegraphics[width=0.95\linewidth]{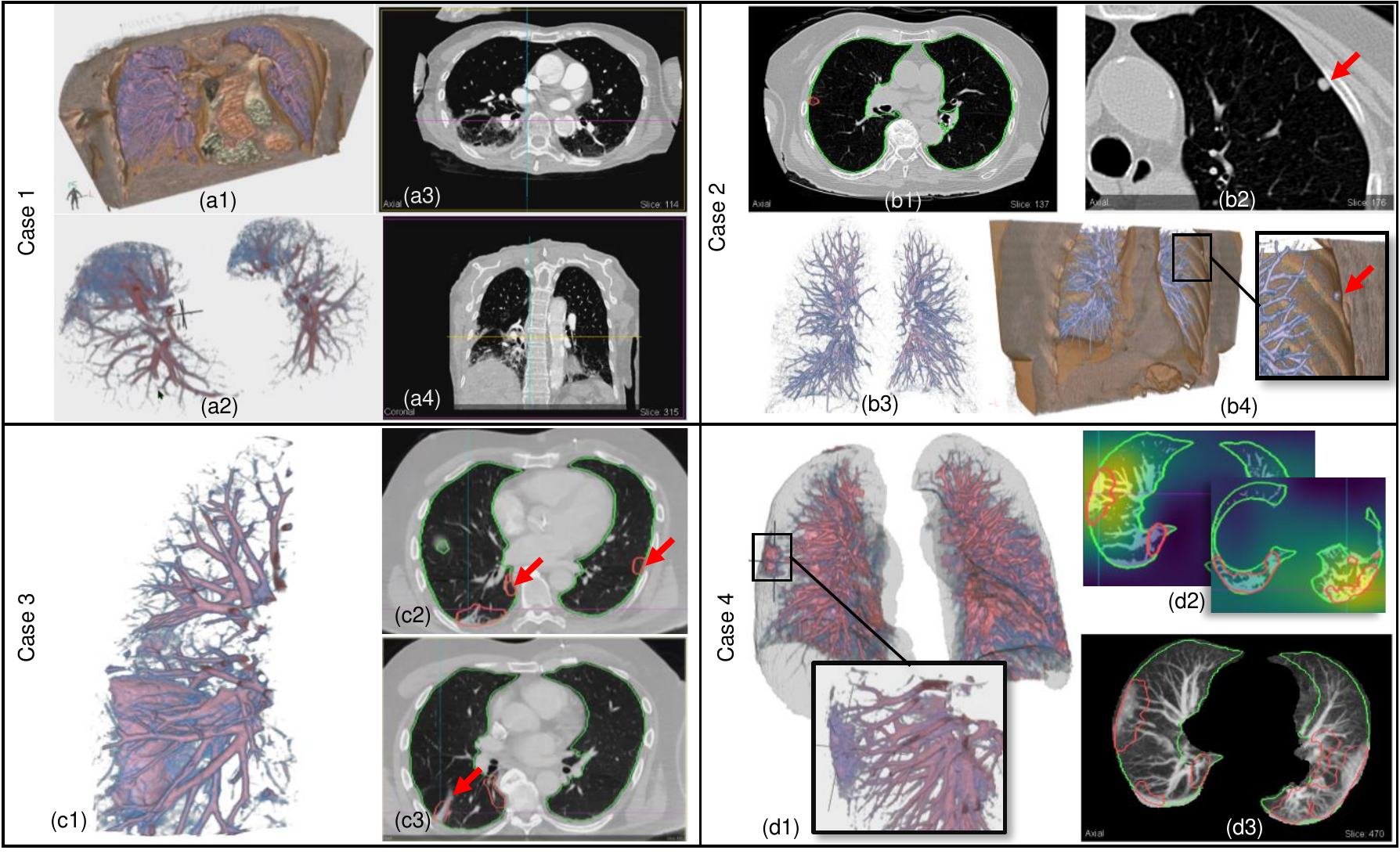}
   \vspace{-1mm} 
    \caption{Case studies performed by our collaborating radiologist (MZ).
    Case 1: (a1) 3D view with lungs and context volume clipped by top and front planes shows affected lower-posterior lungs regions. (a2) Lower posterior lung regions rendered in isolation. (a3) Axial view. (a4) Coronal view showing opacities in the lungs dependent regions.
    Case 2: (b1) Axial view showing very little abnormality detected by lesion segmentation model.
    (b2) Axial view showing nodule (red arrow) identified by radiologist.
    (b3) 3D view of isolated lungs volume.
    (b4) 3D clipped view of lungs and vicinity context volume showing a nodule (red arrow).
    Case 3: (c1) 3D view of isolated lungs showing 3D structure of interlobular wall thickening.
    (c2) Axial view with red arrows showing segmented lesions outlines. The radiologist noticed that these opacities were very subtle and they would have missed them without the outlines drawing attention to them.
    (c3) Axial view with red arrow pointing to the septal thickening shown in 3D view.
    Case 4: (d1) 3D view with lungs outline rendered as translucent surface. The outline provides a good context to understand lesion locations and overall lungs geometry (specially if diseased).
    Zoomed inset shows a closer look at GGO in right lung.
    (d2) Axial view with an overlay of the classification model activation heatmap.
    The model detected COVID-19 on the scan with 99\% certainty.
    (d3) Axial MIP view showing segmented lesion outlines in red.}
    \label{fig:case_studies}
\end{figure*}

\textbf{Case 1:}
The first case analyzed by the radiologist MZ is a 79 year old female who had a chest CT scan with intravenous (IV) contrast (see Fig. \ref{fig:case_studies} Case 1).
MZ inspected the case in both 2D views (particularly using axial slices) and 3D visualization of the lungs.
After glancing at the 3D view of the lungs volume, MZ was quickly able to comment on the distribution of the COVID-19 lesions. MZ pointed out that most lesions proportion were posterior rather than anterior, and were in the lungs dependent region.
This may also happen due to the patient's supine pose as the lungs  may collapse (sub-segmental atelectasis) due to gravity, showing higher opacity in the lungs dependent region regardless of whether the lungs are diseased.
Another case for cause of higher opacities in the lungs posterior region is pulmonary edema, that is, accumulation of fluid in the lungs webbing.
MZ observed an asymmetry in the lesion mass between left and right lungs, which favors an infectious process in the right lower lobe, but was hesitant to strongly classify it as COVID-19 related lesions due to the aforementioned possibility of atelectasis or pulmonary edema. MZ looked at the classification results which confirmed this case as COVID-19 positive with 99\% certainty.
MZ also looked at the activation heatmap which pointed to the same region in the patient's lower right lung which MZ also had identified as possible infectious region.
Indeed, the ground truth label also confirmed that the patient was COVID-19 positive.

\textbf{Case 2:}
The second case is a 77 year old female who had a chest CT without IV contrast (see Fig. \ref{fig:case_studies} Case 2).
MZ identified a pulmonary nodule on the scan. The chest scan did not have any other visible abnormalities. The classification module reported this case as COVID-19 negative, and the lesion localization model also did not identify any abnormalities on the chest volume. MZ was glad to see that the nodule, which is not COVID-19 related, was not misidentified as a COVID-19 related lesion and the classification model was also able to correctly report the case as a negative with 97.8\% certainty.

\textbf{Case 3:}\label{paragraph-case3}
The third case is a 72 year old male who had a chest CT with IV contrast (see Fig. \ref{fig:case_studies} Case 3).
MZ studied the patient through the 2D views and the 3D lungs rendered in isolation with vessels and lesions.
They were able to identify IST in the right lung through the 3D visualizations.
MZ noticed that the lesion localization model showed outlines on the 2D views for subtle abnormal regions that they would not have noticed.
The outlines in the 2D views were able to draw the radiologist's attention to subtle areas of GGOs, which they would have otherwise missed.
MZ explained that early stage COVID-19 lung opacities can be more subtle as they have not fully developed. The outlines in such cases can help in drawing attention to the lesions extent and distribution and support a more thorough examination of the chest CT. A radiology report will often describe the distribution and extent of GGO in terms of how many lobes are covered.

\textbf{Case 4:}
The fourth case is a 77 year old female who had a chest CT without IV contrast (see Fig. \ref{fig:case_studies} Case 4).
MZ looked at the automatic classification results and checked the correlation between the identified lesions and the classification model activation heatmap.
MZ found a decent correlation between the two, but had questions regarding why the lesion localization outlines didn't accurately correspond with the classifier activation heatmap. We explained that the lesion localization model \cite{fan2020inf} was trained on manually segmented lesions and is specifically trained for segmentation tasks, whereas the activation heatmap is part of model results explanation rather than an accurate segmentation.
MZ also expressed the desire to have some control over the heatmap parameters such as the scaling and thresholding parameters of the colormap, which could be used to identify multiple peak points of the heatmap. MZ identified a lesion (GGO) in the right lung that was not particularly highlighted by the heatmap but was identified by the lesion segmentation model (red outlines in 2D views).
The radiologist then looked at the isolated lesion in 3D view along with the lungs surface.
MZ appreciated the 3D view as it was able to provide additional information about the lesion shape and its location, for example, that the lesion is peripheral and provides a general qualitative sense of the lesion size with respect to surrounding features, such as vessels. MZ mentioned that if a radiologist was in doubt about the lesion in 2D views, the 3D views can provide additional qualitative information that may be able to clear up the doubt.
The radiologist also viewed the case with context and lungs rendered in 3D with a coronal clipping plane, as in Fig. \ref{fig:3d_eg}b. MZ explored the visualization with provided transfer function presets. They noticed the heart and the subcutaneous fat region due to the color mapping.
They opined that since COVID-19 diagnosis and the causes of severity of illness are still being investigated and are an active medical research, there is also some interest in correlating body fat with the disease. Visualizing the fat region and perhaps even quantifying it might be a useful for diagnosis and even for research.
Similarly, the rendering of the context volume and use of clipping tool can help visualize the cardiac region. The cardiac region can be of interest for investigating pulmonary embolism, and change in contour of particularly the right heart chamber due to back pressure from the lungs.
This case demonstrates non-dependent areas of opacity (lesions).

\paragraph{}
Beyond the case studies, we had further discussions with both radiologists (MZ and AA). They commented on each of the \sysname\ user interface and visualization component, and their qualitative feedback is summarized below.

\textbf{3D Visualizations:}
Overall, both radiologists were pleased with the 3D visualization capabilities.
MZ mentioned that the vessels in the lungs were clearly visible in 3D.
AA was also able to quickly identify a nodule and interlobular septal wall thickening in the scan using the 3D views as compared to the 2D views.
Both radiologists found that the 3D view of the isolated lungs volume is helpful to quickly identify lesion locations and characterize the distributions (e.g., bilateral/unilateral and location in dependent or non-dependent regions) which is important to identify infectious processes in the lungs. 
MZ mentioned that 3D visualization of blood vessels could also be helpful since non-aerated areas of the lungs get shunted out and blood tends to flow in regions that are more oxygenated. 
The lungs outline rendered as translucent surface provided critical context for understanding lesion distribution, and may also help in visualizing stiffness in the lungs. Stiffness can cause lungs to under-inflate and reduce oxygenation. Identifying these regions can help the physicians for better patient's management in terms of hospitalization, ICU and appropriate ventilation.

\textbf{Clipping Tool:}
3D view of the context volume with clipping planes is robust and provides additional look at the cardiac region and body fat through suitable transfer function presets.
The clipping tool was very useful in localizing 3D ROIs. Both radiologists mentioned that they are familiar with the bounding box based clipping. Such 3D visualizations and interaction tools are not common in their workflow. They are more common to specialized applications such as virtual colonoscopy and virtual bronchoscopy. MZ also mentioned that for COVID-19 diagnosis, virtual endoscopy style navigation would not be useful, and it was better to have an outside rendering of the lungs, such as in \sysname.

\textbf{Transfer Function Design and Presets:}
MZ stated that they prefer different color combinations than the color-maps we had used in the provided presets.
Therefore, they appreciated the facility to allow them to create their own presets and personalize the TFs for different tasks.
AA liked the application of color presets in the 3D clipped (front plane) view, which provided the view point for inspecting the heart region and lungs blood vessels. AA stated that such renderings can be helpful in assessing other conditions, such as pulmonary embolism.

\textbf{MIP views:}
It is a well-known tool to radiologists and many find it very helpful. Particularly, subtle GGOs  and nodules are easily visible in MIP views. AA suggested adding colors, such as application of transfer functions for the MIP views, to further improve its utility.

\textbf{COVID-19 Classifier:}
Both radiologists appreciated that we had incorporated the novel automatic binary classifier into the system,
and that the pre-processing pipeline of the application was entirely automatic. This greatly helps in the utility of the system in the field, since a radiologist is unlikely to spend time interacting with semi-automatic segmentation or classification. MZ noted that it is useful to have a second reader  and the activation heatmap provides explanation for the classifier results and has the potential to improve COVID-19 detection.
They pointed out that the heatmap was generally activated heavily by a singular region rather than being activated by every COVID related lesion. We explained that the activation heatmap may rely heavily on the most prominent lesion and may not get triggered by every lesion.

\textbf{Measurement Tools:}
MZ explained that the 2D measurement tools and automatic lungs and lesions volume measurements are indispensable for assessing the size and growth of lung opacities, and the quantitative assessment of lesion volumes can help in patient management and prognosis.
Furthermore, both AA and MZ mentioned other scenarios where \sysname\  and its visualization capabilities can be useful, for example, to inspect cases of pulmonary embolism and rib fractures in trauma patients as those could be difficult to analyze in 2D axial views.
In addition, we demonstrated \sysname\ to medical trainee JZ to understand the perspective of medical students currently undergoing training for reading CT images.
Based on the automatic classification capability, JZ stated that the \sysname\ could be used as a second reader to the radiologist that can provide additional data points for diagnosis.
JZ also stated that in their personal experience \sysname\ user interface and visualizations appear better than other software applications they have encountered in the hospital during their training,
and that they are interested in using the tool and learning more about its capabilities.
They also suggested that help documentation in the form of tool-tips and question mark buttons should be embedded into the system for easier adoption by new users.

\section{Conclusion and Future Work}\label{sec:conclusion}
We have developed a novel 3D visual diagnosis application, \sysname, for radiological examination of chest CT for suspected COVID-19 patients.
It aims at supporting disease diagnosis and patient management and prognosis decision making.
While lab tests (RT-PCR) are available for screening patients for COVID-19, our \sysname\ visual+DL system
provides a more comprehensive analytical tool for assessing severity and urgency in case of hospitalized patients.
It further supports other analysis tasks (e.g., pulmonary embolism diagnosis) by augmenting the conventional 2D workflow with 3D visualization of not just the lungs but also the context volume of the lungs and heart.
We developed a novel DL classification model for classifying patients as COVID-19 positive/negative, which has high accuracy and reliability. It is integrated into our user interface as a \emph{second reader} along with the visualization system for visual+DL diagnosis. In addition, an activation heatmap generated by our classification model can be overlaid on the 2D views as explainable DL
and as visual+DL decision support.

Diagnosis and treatment of COVID-19 is an active on-going research. We believe that a visual+DL system, such as \sysname\, can play a critical role in not only diagnosing and managing patients but also supporting researchers in further understanding the disease through 3D lung exploration and visualization, 3D lesion morphology, lungs exterior geometry, cardiac region, and full-body scans.
It has been suggested by our expert radiologists to expand our  detection to other forms of abnormalities and diseases, such as cancer lesions, nodules, pulmonary embolism, and other forms of lung fibrosis due to various pneumonia. This would expand our application to a more general chest CT analysis utility, which we plan to pursue in the future.

We further plan to improve our application prototype
by adding an heart segmentation model to support additional analysis of the cardiac region and the heart right chambers, and visualization and quantification of body fat as additional measure. We will also apply other explainable DL methods, such as network dissection methods \cite{netdissect2017}, and compare their performance 
to improve the classification model interpretability. 
We will also explore the possibility of developing automatic viewpoints for structures that are better viewed from specific orientations (e.g., IST).
The data for our study was collected from a single hospital, as is common in initial studies, especially in COVID-19 (e.g., \cite{wang2020weakly}). We will collaborate with other institutions and collect additional data from different populations to perform cross-institution and cross-population validation.
While the incorporated lesion segmentation model by Fan et al.~\cite{fan2020inf} is useful in our current prototype, 
we plan to develop a more accurate segmentation and lesion type classification models.

\acknowledgments{
The chest CT datasets are courtesy of Stony Brook University Hospital. We thank Luke Cesario and Michael Yao for coding help and to Dr. Almas Abbasi and Joshua Zhu for application evaluation and qualitative feedback.
This work was funded in part by NSF grants CNS1650499, OAC1919752, ICER1940302, and IIS2107224, and OVPR-IEDM COVID-19 grant.
}

\bibliographystyle{abbrv-doi}

\bibliography{references}
\end{document}